\documentclass[aps,prb,floatfix,twocolumn,footinbib,superscriptaddress]{revtex4-1}
\usepackage{epsfig}
\usepackage{graphicx}
\usepackage{dcolumn}   
\usepackage{bm}
\usepackage{amssymb}
\usepackage{mathrsfs}
\usepackage{amsmath}
\usepackage{bbold}
\usepackage{epstopdf}

 \usepackage[colorlinks=true,linkcolor=blue,anchorcolor=blue,citecolor=blue,urlcolor=blue]{hyperref}

\begin{document}

\title{Energy transfer between two vacuum-gapped metal plates: Coulomb fluctuation and electron tunneling}

\preprint{APS/123-QED}
\author{Zu-Quan Zhang}
\author{Jing-Tao L\"u}
\email{jtlu@hust.edu.cn}
\affiliation{School of Physics and Wuhan National High Magnetic Field Center, Huazhong University of Science and Technology, 430074 Wuhan, P. R. China}
\author{Jian-Sheng Wang}
\email{phywjs@nus.edu.sg}
\affiliation{Department of Physics, National University of Singapore, Singapore 117551, Republic of Singapore}

\date{\today}

\begin{abstract}
Recent experimental measurements for near-field radiative heat transfer between two bodies have been able to approach the gap distance within $2 \; \textrm{nm}$, where the contributions of Coulomb fluctuation and electron's tunneling are comparable. Using the nonequilibrium Green's function method in the $G_{0}W_{0}$ approximation, based on a tight-binding model, we obtain for the energy current a Caroli formula from the Meir-Wingreen formula in the local equilibrium approximation. Also, the Caroli formula is consistent with the evanescent part of the heat transfer from the theory of fluctuational electrodynamics. We go beyond the local equilibrium approximation to study the energy transfer in the crossover region from electron tunneling to Coulomb fluctuation based on a numerical calculation.
\end{abstract}

\maketitle

\section{\label{sec:Intro} Introduction}
Recently, heat transfer through a gap of nanometer scale between two bodies has attracted enormous interest~\cite{Volokitin_RMP2007, SongBai2015}. Researchers have made great efforts~\cite{Kittel2005, Shen2009, Rousseau2009, Altfeder2010, Kim2015, Song2016, Kloppstech2017, Cui_NatCom2017} into the measurement of heat transfer for reduced gap size in the near-field region (gap size smaller than the Wien's wavelength), from thousands of nanometers to a few nanometers. This is inspired by the previous interesting discovery~\cite{hargreaves1969, Polder1971} in the 70s that the radiative heat transfer in the near-field distance was much larger than that of the far field value predicted by Planck's law of black-body radiation. The very recent experiments~\cite{Kim2015, Song2016, Cui_NatCom2017} were conducted at the extreme near-field distances of less than 10 $\textrm{nm}$. Good agreement was found between the values predicted by the conventional theory of fluctuational electrodynamics~\cite{rytov1959theory, Polder1971, Rodriguez2013} and the experimental results.  Heat transfer of about orders of magnitude larger than that of the black-body limit was explained as a result of the surface phonon polaritons for dielectric material or the normal evanescent modes for the metal.

For gap distances within a few nanometers, several experiments~\cite{Kittel2005, Altfeder2010, Worbes2013, Kloppstech2017} found much larger magnitude of heat transfer than the value predicted by the fluctuational electrodynamics.  On the one hand, different theoretical models have been proposed to explain these experiments, such as calling for microscopic theories  beyond the macroscopic fluctuational electrodynamics, or different physical mechanism of phonon tunneling rather than energy mediated by electromagnetic field. On the other hand, a very recent experiment~\cite{Cui_NatCom2017} showed that the experimental results were very sensitive to the condition of the vacuum-gap, i.e., potential contaminants could lead to much higher thermal conductance than that predicted by fluctuational electrodynamics. The experiment showed good support for the fluctuational electrodynamics without the contaminants.

The theory of fluctuational electrodynamics describes the radiation of the electromagnetic field by the thermal current fluctuations, which are related to the dielectric function of the system by the fluctuation-dissipation relation in thermal equilibrium condition. The heat transfer given by the theory of fluctuational electrodynamics has an evanescent part and a propagating part. From a different point of view, the Coulomb interaction from charge fluctuations can transfer energy through a vacuum gap and it is rarely studied~\cite{Yu2017, Mahan2017, Wang2017, Jiang2017}. Based on a microscopic quantum mechanical model for describing heat tunneling between two metals, Mahan~\cite{Mahan2017} found that the electron Coulomb interaction had a dominant contribution to heat transfer at small gaps. Yu~\cite{Yu2017} studied the contribution of Coulomb fluctuation to heat transfer using a quantum mechanical linear response theory. Our previous work~\cite{Jiang2017} gave a Caroli formula~\cite{Caroli1971} for studying this problem, and it is shown that the contribution of heat transfer due to Coulomb interaction corresponds to the evanescent part given by the theory of fluctuational electrodynamics. It can be shown that these results are consistent despite of using different methods.

In this work, based on a tight-binding model on a cubic lattice, we study the contribution of Coulomb fluctuation to energy transfer between two separated metals using the nonequilibrium Green's function (NEGF) method~\cite{Wang2008, Wang2014} based on the random phase approximation (RPA). Firstly, when the electron's tunneling through the vacuum gap is not included, we find that by using the local equilibrium approximation, the Meir-Wingreen formula using $G_{0}W_{0}$ approximation for calculating the energy transfer reduces to the Caroli formula, and it is consistent with previous results~\cite{Yu2017, Mahan2017}.  More importantly, by allowing electrons to tunnel through the vacuum gap, we study the crossover of energy transfer from conducting limit to Coulomb fluctuation limit. We find that electron tunneling can drastically enhance the heat transfer efficiency. We also discuss the influence of tunneling barrier on the heat transfer. This is relevant to  the experimental case~\cite{Cui_NatCom2017} where the energy transfer at extreme near distance is very sensitive to contaminants, which can modify the tunneling barrier.

\section{\label{sec:ModelAndMethod} Model and Method}
\subsection{Model}
We deal with the Coulomb interaction by the scalar photons in the framework of classical electromagnetic field, which was used recently to the quantum dot model and the graphene case\cite{Wang2017, Jiang2017}. The Lagrangian density is given in the Lorentz gauge condition by\cite{cohen1989photons}
\begin{equation} \label{eq:Lagrangian-EM}
\mathcal{L} = \frac{\epsilon_{0}}{2} \left[ { \dot{ \bm{A} } }^2 - c^2 \sum_{ij} ( \partial_{i} A_{j} )^2 - \left( \frac{ \dot{\phi} }{c} \right)^2  + ( \nabla \phi )^2  \right] - \rho \phi + \bm{j} \bm{\cdot} \bm{A},
\end{equation}
where  $\epsilon_{0}$ is the vacuum permittivity, $c$ is the constant velocity of light in the vacuum, $\bm {A}$ is the vector potential, and $\phi$ is the scalar potential, $\rho$ is the charge density, $\bm{j}$ is the current density. Using the Euler-Lagrange equation, we get the equation of motion for the scalar field
\begin{equation} \label{eq:EQM}
\nabla^2 \phi - \frac{1}{c^2} \frac{\partial^2 \phi}{\partial t^2} = - \frac{\rho}{\epsilon_0}.
\end{equation}
Canonical momentum of the scalar potential is obtained by
\begin{equation} \label{eq:CanonicalMomentum}
\bm{\pi} = \frac{\partial \mathcal{L}}{ \partial \dot{\phi} } = - \frac{\epsilon_0}{c^2} \dot{\phi},
\end{equation}
and the commutation relation is
\begin{equation} \label{eq:CommutationRelation}
 [ \phi (\bm{r}), \bm{\pi} (\bm{r'}) ] = i \hbar \delta ( \bm{r} - \bm{r'} ) .
\end{equation}
We restrict our discussion to the scalar photons and omit the vector potential in the following text. The Hamiltonian density of the scalar field is given by
\begin{equation} \label{eq:Hphi}
\mathcal{H_{\phi}} = \bm{\pi} \dot{\phi} - \mathcal{L} = -\frac{\epsilon_0}{2} \left[ ( \frac{\dot{\phi}}{c} )    ^2 + ( \nabla \phi )^2  \right] + \rho \phi.
\end{equation}

\begin{figure}
\centering
\includegraphics[width=8.5 cm]{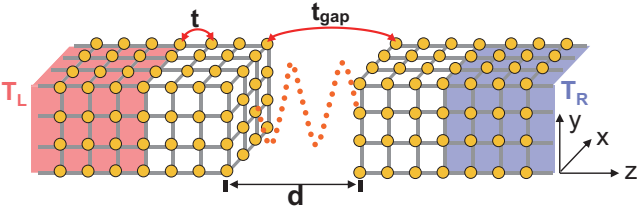}
\caption{Tight-binding model for energy transfer between two vacuum-gapped semi-infinite cubic lattices by Coulomb fluctuation or electron tunneling.}
\label{fig:Model-cubic}
\end{figure}

A tight-binding model of two parallel aligned semi-infinite cubic lattices is used to describe the vacuum-gapped metal plates, as shown in Fig.~\ref{fig:Model-cubic}. Hamiltonian for the noninteracting electrons is
\begin{equation} \label{eq:He}
H_{e} = \sum_{\langle ij\rangle} t_{ij} c^{\dag}_{i} c_{j},
\end{equation}
where $\langle ij\rangle$ denotes the nearest neighbor pairs by site $i$ and $j$. As shown in Fig.~\ref{fig:Model-cubic}, we set all the nearest neighbor hopping parameters in the cubic lattices to be $t$. Electron's tunneling through the vacuum gap is described by the hopping parameter $t_{\textrm{gap}}$. $c_{i}^{\dag}$ ($c_{i}$) is the electron's creation (annihilation) operator at site $i$. In this case, the charge density in Eq.~(\ref{eq:Hphi}) is $ \rho(\bm{r}) =- e \sum_i( c_{i}^{\dag} c_{i} - n_{\textrm{ion}}) \delta(\bm{r} - \bm{r}_i) $, where $-e$ is the electron's charge and $n_{\textrm{ion}}$ is the number of ion's charge per site of the positive background. Hamiltonian for the scalar field can be written as
\begin{equation} \label{eq:Hamil-Scalar}
H_{\phi} = \int d^3 \bm{r} \mathcal{H}_{\phi} = H_{\phi}^{0} + H_{\textrm{int}},
\end{equation}
Here $H_{\phi}^{0} = - \frac{\epsilon_{0}}{2} \int d^3 \bm{r} \left[ ( \frac{\dot{\phi}}{c} )^2 + ( \nabla \phi )^2  \right] $ is the Hamiltonian for the scalar photons in free space, and $H_{\textrm{int}} = -e \sum_{i} ( c_{i}^{\dag} c_{i} - n_{\textrm{ion}})  \phi_{i}$ is the Hamiltonian for the interaction between the scalar photons and the charges. The total Hamiltonian for our system is given by
\begin{equation} \label{eq:Htot}
H = H_{e} + H_{\phi}^{0} + H_{\textrm{int}}.
\end{equation}

It is convenient to write the Hamiltonian in $\bm{k}$ space for the $x$-$y$ direction.  The electron's Hamiltonian in Eq.~(\ref{eq:He}) can be written as
\begin{equation} \label{eq:He-kspace}
H_{e} = - \sum_{n} \sum_{\bm{k}} t f(\bm{k}) c^{\dag}_{n \bm{k}} c_{n \bm{k}} - \sum_{\langle nm\rangle} \sum_{\bm{k}} t_{mn} c_{m \bm{k}}^{\dag} c_{n \bm{k}},
\end{equation}
where $f(\bm{k}) = 2 \cos(k_{x} a) + 2 \cos(k_{y} a) $, with $a$ being the lattice constant, and $m$, $n$ are the layer indices along the $z$ direction.  Hamiltonian of the interacting part in Eq.~(\ref{eq:Hamil-Scalar}) can be written as
\begin{equation} \label{eq:Hint-kspace}
\begin{split}
H_{\textrm{int}} =& - e \frac{1}{\sqrt{N}} \sum_{n} \sum_{\bm{k}, \bm{q}} c_{n,\bm{k}}^{\dag} c_{n, \bm{k} - \bm{q}} \phi_{n, \bm{q}}  \\
&+  e n_{\textrm{ion}} \frac{1}{\sqrt{N}} \sum_{n, \bm{k}} \phi_{n}(\bm{q}=0),
\end{split}
\end{equation}
where $N$ is the number of unit cells in the $x$-$y$ plane.
In the following text, we restrict our discussion to the charge-neutrality case $ \sum_{\bm{k}} ( c_{n \bm{k}}^{\dag} c_{n \bm{k}} - n_{\textrm{ion}} ) = 0$, so that the combination of the Hartree term\cite{combescot1993hartree, PhysRevB.53.3861} ($\bm{q}=0$) and the positive background in Eq.~(\ref{eq:Hint-kspace}) is 0.

\subsection{Method}
We use the NEGF method to study the transport properties of the systems. The Green's functions (GFs) for the scalar photons and electrons are defined respectively as
\begin{equation} \label{DefG-photon}
D(\bm{r},\tau ; \bm{r}', \tau') = - \frac{i}{\hbar} \langle \mathcal{T} \phi^{H}(\bm{r},\tau) \phi^{H}(\bm    {r'},\tau') \rangle,
\end{equation}
and
\begin{equation} \label{DefG-electron}
G(\bm{r},\tau; \bm{r}', \tau') = - \frac{i}{\hbar} \langle \mathcal{T} c^{H}(\bm{r},\tau) c^{\dag, H}(\bm    {r'},\tau') \rangle,
\end{equation}
where  $\mathcal{T}$ is the time order operator on the Keldysh contour. $\phi^{H}(\bm{r},\tau)$ is the operator for the scalar photons in the Heisenberg representation, and $c^{H}(\bm{r},\tau)$ is the electron's annihilation operator in real space. Knowing the equation of motion and the commutation relation for the scalar photons, given by Eq.~(\ref{eq:EQM}) and (\ref{eq:CommutationRelation}) respectively, we can get the equation of motion for the GF of scalar photons in free space as
\begin{equation} \label{eq:EQM-D0GF}
\left( \frac{1}{c^2} \frac{ \mathrm{d}^2 }{ \mathrm{d} t^2 } - \nabla^2 \right) D^{0} (\bm{r},t ; \bm{    r'}, t') = \frac{1}{\epsilon_0} \delta(\bm{r} - \bm{r'}) \delta(t - t'),
\end{equation}
which is just the D'Alembert equation. The retarded GF can be solved from Eq.~(\ref{eq:EQM-D0GF}) in frequency domain with the $x$-$y$ direction in $\bm{q}$ space to adapt to the shape of our system,
\begin{equation} \label{eq:Dreen0_R}
D^{0,r} (\bm{q}, z-z', \omega) = \frac{i e^{i \sqrt{ (\omega + i \eta)^2 / c^2 - {\bm{q}}^2 }  |z-z'|} }{2 s_0 \epsilon_0 \sqrt{ (\omega + i \eta)^2 / c^2 - {\bm{q}}^2 } },
\end{equation}
where $s_{0} = a^2$ is the area of the unit cell in the $x$-$y$ direction, and $\eta$ is a positive infinitesimal number. Since the size of our system is very small compared with the light velocity, we neglect the retardation by making $c \rightarrow \infty$ in the quasi-static limit. This is necessary to meet the Lorentz gauge condition in the absence of vector potential in our case. Then the GF in Eq.~(\ref{eq:Dreen0_R}) becomes
\begin{equation} \label{eq:Dreen0_R_inf}
D^{0,r} (\bm{q}, z-z') = \frac{ e^{- q |z-z'|} }{2 s_0 \epsilon_{0} q}.
\end{equation}

\begin{figure}
\centering
\includegraphics[width=8.5 cm]{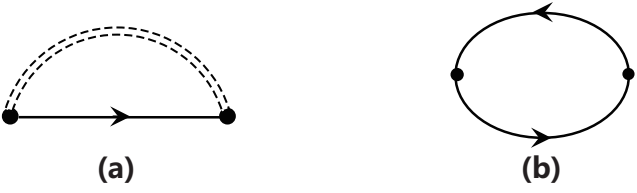}
\caption{(a) Diagram representation of the Fock term for the $G_{0}W_{0}$ approximation in calculating the self-energy of the electrons. (b) The bubble diagram shows that RPA is used for the self-energy of the scalar photons.}
\label{fig:FockBubble}
\end{figure}

The interaction between the scalar photons and the electrons is very similar to that of the electron-phonon interaction. For the latter it can be solved by using the self-consistent Born approximation (SCBA) with the self-energy given by the Hartree-Fock and polarization terms~\cite{HYLDGAARD19941, PhysRevB.75.205413, PhysRevB.76.165418}. For the former, if we also want to solve it by using the SCBA in the same way, we find that this is the case of the self-consistent $GW$ ($\textrm{sc}GW$) approximation~\cite{PhysRev.139.A796} in dealing with electron-electron interaction~\cite{PhysRevB.54.8411}.
Some proposals have been made~\cite{PhysRevLett.96.226402, PhysRevLett.83.241,PhysRevB.84.195114} and it is still under development~\cite{PhysRevB.89.155417,PhysRevB.95.195131}. We restrict our discussion to the simpler $G_{0}W_{0}$ approximation in this text. The RPA~\cite{bruus2004many} is used for the screening of the scalar photons. Diagram representations for the $G_{0}W_{0}$ approximation with the RPA are shown in Fig.~\ref{fig:FockBubble}.

Firstly, we solve the electron's GF without the interaction between the electron and the scalar field. The retarded GF is given by $\bm{G}^{0,r}(\bm{k},E) = [ (E + i \eta) \mathbb{1} - \bm{H}_{e}^{C}(\bm{k},E)  - \bm{\Sigma}_{\rm{leads}}^{r}(\bm{k},E)]^{-1}$, where $\mathbb{1}$ is the identity matrix. $\bm{H}_{e}^{C}(\bm{k},E)$ is the Hamiltonian matrix from Eq.~(\ref{eq:He-kspace}) in the central region without the interaction between the electron and the scalar photons. $\bm{\Sigma}_{\rm{leads}}^{r}= \bm{\Sigma}_{L}^{r} + \bm{\Sigma}_{R}^{r}$ is the self-energy for the two noninteracting cubic electrodes, with $\bm{\Sigma}_{L(R)}^{r} = t^2 \bm{g}_{L(R)}^{r}$,  and $\bm{g}_{L(R)}^{r}$ is the surface GF for the leads. The lesser GF is obtained by the Keldysh equation $\bm{G}^{0,<} = \bm{G}^{0,r} (\bm{\Sigma}_{L}^{<} + \bm{\Sigma}_{R}^{<}) \bm{G}^{0,a}$ for steady state transport, with $\bm{\Sigma}_{\alpha}^{<} = - f_{\alpha}(\bm{\Sigma}_{\alpha}^{r}- \bm{\Sigma}_{\alpha}^{a})$, $\alpha = L, R$. Secondly, we use the Dyson equation to solve the full GFs, where the interaction between the electrons and the scalar field is included as a perturbation to the noninteracting GFs.  For electrons, we have $\bm{G}^{r}(\bm{k},E) = \bm{G}^{0,r}(\bm{k},E) + \bm{G}^{0,r}(\bm{k},E) \bm{\Sigma}^{F,r}(\bm{k},E)  \bm{G}^{r}(\bm{k},E)$, and $\bm{G}^{<}= \bm{G}^{r} \bm{\Sigma}_{\rm{tot}}^{<} \bm{G}^{a}$. Here $\bm{\Sigma}_{\rm{tot}}^{<} = \bm{\Sigma}_{\rm{leads}}^{<} + \bm{\Sigma}^{F,<}$ is the total self-energy. For scalar photons, we have $\bm{D}^{r}(\bm{q},\omega) = \bm{D}^{0,r}(\bm{q}) + \bm{D}^{0,r}(\bm{q}) \bm{\Pi}^{0,r}(\bm{q},\omega)  \bm{D}^{r}(\bm{q},\omega)$, and $\bm{D}^{<}= \bm{D}^{r} \bm{\Pi}^{0,<} \bm{D}^{a}$. Diagram representations for the self-energies are shown in Fig.~\ref{fig:FockBubble}. The Fock self-energies of the electrons are
\begin{equation}  \label{eq:Fock-lesser}
\begin{split}
\Sigma_{mn}^{F,<}(\bm{k},E) =&i \hbar e^2 \frac{1}{N} \sum_{\bm{q}} \int \frac{\textrm{d} \omega}{2 \pi} G_{mn}^{0,<}(\bm{k} - \bm{q}, E - \hbar \omega)  \\
& \times D_{mn}^{<}(\bm{q},\omega),
\end{split}
\end{equation}
and
\begin{equation}  \label{eq:Fock-retarded}
\begin{split}
\Sigma_{mn}^{F,r}(\bm{k},E) =& i \hbar e^2 \frac{1}{N} \sum_{\bm{q}} \int \frac{\textrm{d} \omega}{2 \pi} \Big{\{} \big{[} G_{mn}^{0,r}(\bm{k} - \bm{q}, E - \hbar \omega)  \\
& + G_{mn}^{0,<}(\bm{k} - \bm{q}, E - \hbar \omega) \big{]}D_{mn}^{r}(\bm{q},\omega)  \\
& + G_{mn}^{0,r}(\bm{k} - \bm{q}, E - \hbar \omega) D_{mn}^{<}(\bm{q},\omega) \Big{\}} ,
\end{split}
\end{equation}
where the subscript $m,n$ are for the layer indices along the $z$ direction, for example $D_{mn}^{r}(\bm{q},\omega) \equiv D^{r}(\bm{q}, z_{m},z_{n} ,\omega) $. The polarization self-energies are
\begin{equation}
\begin{split}
\Pi_{mn}^{0,<} (\bm{q},\omega) =& - i \hbar e^2 \frac{1}{N} \sum_{\bm{k}} \int \frac{\textrm{d} E}{2 \pi \hbar} G_{mn}^{0,<}(\bm{k},E) \\
& \times G_{nm}^{0,>}(\bm{k} - \bm{q}, E - \hbar \omega),
\end{split}
\end{equation}
and
\begin{equation}
\begin{split}
\Pi_{mn}^{0,r} (\bm{q},\omega) =& - i \hbar e^2 \frac{1}{N} \sum_{\bm{k}} \int \frac{\textrm{d} E}{2 \pi \hbar} \big{[} G_{mn}^{0,r}(\bm{k},E) \\
& \times G_{nm}^{0,<}(\bm{k} - \bm{q}, E - \hbar \omega) + G_{mn}^{0,<}(\bm{k},E)  \\
& \times G_{nm}^{0,a}(\bm{k} - \bm{q}, E - \hbar \omega) \big{]}.
\end{split}
\end{equation}
The energy current per unit area flowing out of lead $\alpha$ is given by the Meir-Wingreen formula~\cite{PhysRevLett.68.2512, PhysRevB.50.5528} as
\begin{equation}  \label{eq:MWformula}
\begin{split}
P_{\alpha} =& \frac{1}{A} \int \frac{\textrm{d} E}{2 \pi \hbar} E \sum_{\bm{k}} \textrm{Tr} \big[   \bm{\Sigma}_{\alpha}^{<}(\bm{k},E) \bm{G}^{>}(\bm{k},E) \\
& - \bm{\Sigma}_{\alpha}^{>}(\bm{k},E)  \bm{G}^{<}(\bm{k},E) \big],
\end{split}
\end{equation}
where $A$ is the area of the metal plate.

\section{\label{sec:CalAndResult} Caroli formula}
\begin{figure}
\centering
\includegraphics[width=8.5 cm]{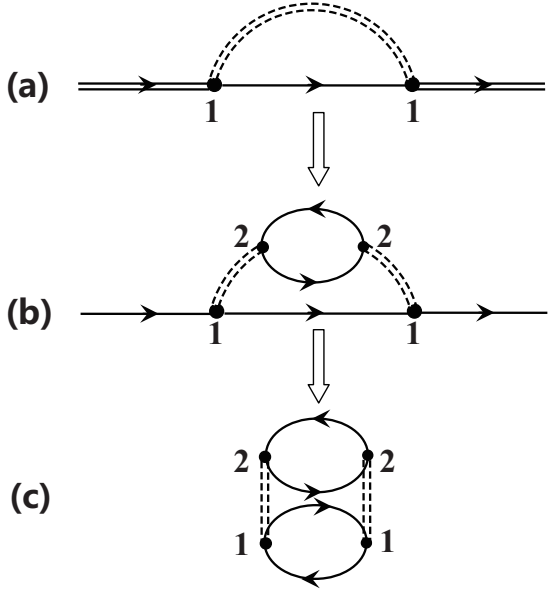}
\caption{Illustrative diagrams for the derivation from Meir-Wingreen formula to Caroli formula. (a) Keldysh  equation is used for the electron's GF in the $G_{0}W_{0}$ approximation in Eq.~(\ref{eq:PL-2}). (b) Keldysh equation is used for the scalar photon's GF using the RPA in Eq.~(\ref{eq:DlRPA}). (c) It is suggestive to close the lower bubble diagram from a symmetric point of view. This is shown from Eq.~(\ref{eq:PLA}) to Eq.~(\ref{eq:PLA2}).}
 \label{fig:MW2Caroli}
\end{figure}

Here we give the relation between the Caroli formula and the Meir-Wingreen formula by using the $G_{0}W_{0}$ approximation and the local equilibrium approximation in the case of energy transfer by Coulomb fluctuation through a vacuum gap. The electron's tunneling through the vacuum gap is neglected by making $t_{\textrm{gap}}$=0, so that the electron's GFs connecting the left and right interacting layers are zero, for example $\bm{G}_{\bm{1}\bm{2}}^{r}(E) = 0$, where $\bm{1}$($\bm{2}$) denotes the left (right) interacting layers. From the Meir-Wingreen formula, the energy flow out of the left lead becomes
\begin{equation} \label{eq:PL-1}
\begin{split}
P_{L} = & \frac{1}{A} \int \frac{\textrm{d} E}{2 \pi \hbar} E \textrm{Tr} \big{[} \bm{\Sigma}_{L}^{<}(E)  \bm{G}_{\bm{1}\bm{1}}^{r}(E)  \bm{\Sigma}_{\bm{1}\bm{1}}^{>}(E) \bm{G}_{\bm{1}\bm{1}}^{a}(E)  \\
& - \bm{\Sigma}_{L}^{>}(E)  \bm{G}_{\bm{1}\bm{1}}^{r}(E)  \bm{\Sigma}_{\bm{1}\bm{1}}^{<}(E) \bm{G}_{\bm{1}\bm{1}}^{a}(E) \big{]},
\end{split}
\end{equation}
where $\bm{\Sigma}_{\bm{1}\bm{1}}^{<}(E) = \bm{\Sigma}_{\bm{1}\bm{1}}^{F,<}(E) + \bm{\Sigma}_{L}^{<}(E) $ is the total self-energy. Neglecting $\bm{\Sigma}_{L}^{<}(E)$ in the self-energy which does not contribute to energy transfer, we get from Eq.~(\ref{eq:PL-1})
\begin{equation} \label{eq:PL-2}
\begin{split}
P_{L} = & \frac{1}{A} \int \frac{\textrm{d} E}{2 \pi \hbar} E \textrm{Tr}\{ \bm{\Sigma}_{L}^{<}(E)  \bm{G}_{\bm{1}\bm{1}}^{r}(E)  \bm{\Sigma}_{\bm{1}\bm{1}}^{F,>}(E) \bm{G}_{\bm{1}\bm{1}}^{a}(E)  \\
& - \bm{\Sigma}_{L}^{>}(E)  \bm{G}_{\bm{1}\bm{1}}^{r}(E)  \bm{\Sigma}_{\bm{1}\bm{1}}^{F,<}(E) \bm{G}_{\bm{1}\bm{1}}^{a}(E) \}.
\end{split}
\end{equation}
The energy transfer mediated by scalar field from the left interacting layers to the right interacting layers is reflected in the Fock self-energy from Eq.~(\ref{eq:PL-2}) by the first term of the following Keldysh equation
\begin{equation} \label{eq:DlRPA}
\begin{split}
\bm{D}_{\bm{1}\bm{1}}^{<}(\omega) =& \bm{D}_{\bm{1}\bm{2}}^{r}(\omega)  \bm{\Pi}_{\bm{2}\bm{2}}^{0,<} (\omega)     \bm{D}_{\bm{2}\bm{1}}^{a}(\omega)  \\
&+ \bm{D}_{\bm{1}\bm{1}}^{r}(\omega) \bm{\Pi}_{\bm{1}\bm{1}}^{0,<}(\omega) \bm{D}_{\bm{1}\bm{1}}^{a}(\omega) ,
\end{split}
\end{equation}
where the second term is related to the energy circulation inside the left interacting layers and it doesn't contribute to energy transfer. After getting the energy transfer process, we can replace the full GFs of the electrons in Eq.~(\ref{eq:PL-2}) by the noninteracting ones as a lowest order approximation. We have
\begin{widetext}
\begin{eqnarray}
P_{L}^{A} \approx  \textrm{Tr} \Bigg{\{} \frac{1}{A} \int \frac{\textrm{d} E}{2 \pi \hbar} E \bm{\Sigma}_{L}^{<}(E)  \bm{G}_{\bm{1}\bm{1}}^{0,r}(E)  (i \hbar e^2) \int \frac{\textrm{d} \omega}{2 \pi} \Big{\{} \bm{G}_{\bm{1}\bm{1}}^{0,>}(E - \hbar \omega) \circ \big{[} \bm{D}_{\bm{1}\bm{2}}^{r}(\omega)  \bm{\Pi}_{\bm{2}\bm{2}}^{0,>}(\omega) \bm{D}_{\bm{2}\bm{1}}^{a}(\omega) \big{]} \Big{\}}  \bm{G}_{\bm{1}\bm{1}}^{0,a}(E) 
 \Bigg{\}},  \label{eq:PLA}  \\
P_{L}^{B} \approx - \textrm{Tr} \Bigg{\{} \frac{1}{A} \int \frac{\textrm{d} E}{2 \pi \hbar} E \bm{\Sigma}_{L}^{>}(E)  \bm{G}_{\bm{1}\bm{1}}^{0,r}(E)  (i \hbar e^2) \int \frac{\textrm{d} \omega}{2 \pi} \Big{\{} \bm{G}_{\bm{1}\bm{1}}^{0,<}(E - \hbar \omega) \circ \big{[} \bm{D}_{\bm{1}\bm{2}}^{r}(\omega)  \bm{\Pi}_{\bm{2}\bm{2}}^{0,<}(\omega) \bm{D}_{\bm{2}\bm{1}}^{a}(\omega) \big{]} \Big{\}} \bm{G}_{\bm{1}\bm{1}}^{0,a}(E)
 \Bigg{\}}, \label{eq:PLB}
\end{eqnarray}
\end{widetext}
and $P_{L} = P_{L}^{A} + P_{L}^{B}$. Here we have defined the matrix notation $\bm{A} \circ \bm{B} \equiv A_{ij} B_{ij}$ for writing the Fock self-energy. Making $E' = E - \hbar \omega$, and using $\bm{G}_{\bm{1}\bm{1}}^{0,<(>)}(E) =\bm{G}_{\bm{1}\bm{1}}^{0,a}(E) \bm{\Sigma}_{L}^{<(>)}(E) \bm{G}_{\bm{1}\bm{1}}^{0,r}(E) $ because the bare GFs are in local equilibrium neglecting the electron's tunneling, we get from Eq.~(\ref{eq:PLA})
\begin{widetext}
\begin{equation}   \label{eq:PLA-1}
\begin{split}
P_{L}^{A} =& \textrm{Tr} \Bigg{\{} \frac{1}{A} \int \frac{\textrm{d} E'}{2 \pi \hbar} \int \frac{\textrm{d} \omega}{2 \pi}   E' (i \hbar e^2) \bm{G}_{\bm{1}\bm{1}}^{0,<}(E' + \hbar \omega) \Big{\{}  \bm{G}_{\bm{1}\bm{1}}^{0,>}(E') \circ \big{[} \bm{D}_{\bm{1}\bm{2}}^{r}(\omega)  \bm{\Pi}_{\bm{2}\bm{2}}^{0,>}(\omega) \bm{D}_{\bm{2}\bm{1}}^{a}(\omega) \big{]} \Big{\}}
 \Bigg{\}}  \\
& + \textrm{Tr} \Bigg{\{} \frac{1}{A} \int \frac{\textrm{d} E'}{2 \pi \hbar} \int \frac{\textrm{d} \omega}{2 \pi}   \hbar \omega (i \hbar e^2) \bm{G}_{\bm{1}\bm{1}}^{0,<}(E' + \hbar \omega) \Big{\{} \bm{G}_{\bm{1}\bm{1}}^{0,>}(E') \circ \big{[} \bm{D}_{\bm{1}\bm{2}}^{r}(\omega)  \bm{\Pi}_{\bm{2}\bm{2}}^{0,>}(\omega) \bm{D}_{\bm{2}\bm{1}}^{a}(\omega) \big{]} \Big{\}}
 \Bigg{\}}.
\end{split}
\end{equation}
\end{widetext}
Making $\omega' = - \omega $ for the first term in Eq.~(\ref{eq:PLA-1}), and using the relations $\bm{D}^{r(a)}(-\omega) = [\bm{D}^{r(a)}(\omega)]^{\ast}$, $\bm{\Pi}^{0,<}(- \omega) = \bm{\Pi}^{0,>}(\omega)$, we get from Eq.~(\ref{eq:PLA-1}) $P_L^{A} = P_L^{A1} + P_L^{A2}$, $P_L^{A1} = - P_{L}^{B}$, with
\begin{equation}   \label{eq:PLA2}
P_L^{A2}= - \int \frac{\textrm{d} \omega}{2 \pi  A} \hbar \omega \textrm{Tr} [ \bm{\Pi}_{\bm{1}\bm{1}}^{0,<}(\omega)  \bm{D}_{\bm{1}\bm{2}}^{r}(\omega)   \bm{\Pi}_{\bm{2}\bm{2}}^{0,>}(\omega)  \bm{D}_{\bm{2}\bm{1}}^{a}(\omega) ].
\end{equation}
We have $P_{L} = P_{L}^{A2}$. Similarly we can get $P_{L} = P_{L}^{B2}$ from Eq.~(\ref{eq:PLB}), with
\begin{equation}   \label{eq:PLB2}
P_L^{B2} =  \int \frac{\textrm{d} \omega}{2 \pi A} \hbar \omega \textrm{Tr}[ \bm{\Pi}_{\bm{1}\bm{1}}^{0,>}(\omega)  \bm{D}_{\bm{1}\bm{2}}^{r}(\omega)   \bm{\Pi}_{\bm{2}\bm{2}}^{0,<}(\omega)  \bm{D}_{\bm{2}\bm{1}}^{a}(\omega) ].
\end{equation}
The polarization self-energy obeys the Bose distribution in the local equilibrium condition, such as $\bm{\Pi}_{\bm{1}\bm{1}}^{0,<}(\omega) = i 2 N_{B}^{L}(\omega) \textrm{Im}[\bm{\Pi}_{\bm{1}\bm{1}}^{0,r}(\omega)]$. $N_{B}^{L}(\omega)$ is the Bose distribution function for the temperature of the left lead. Using $P_{L} = \frac{1}{2} (P_{L}^{A2} + P_{L}^{B2})$, we get a Landauer form for the energy current density
\begin{equation}  \label{eq:Caroli}
P_L = \frac{1}{A} \int_{0}^{\infty} \frac{\textrm{d} \omega}{2 \pi} \hbar \omega [N_{B}^{L}(\omega) - N_{B}^{R}(\omega)]   S(\omega),
\end{equation}
where the transmission function is given by the Caroli formula
\begin{equation} \label{eq:Trans}
 S(\omega) = 4 \textrm{Tr} \big{\{} \textrm{Im}[ \bm{\Pi}_{\bm{1}\bm{1}}^{0,r}(\omega) ] \bm{D}_{\bm{1}\bm{2}}^{r}(\omega) \textrm{Im}[ \bm{\Pi}_{\bm{2}\bm{2}}^{0,r}(\omega)]  \bm{D}_{\bm{2}\bm{1}}^{a}(\omega) \big{\}}.
 \end{equation}
 We show in Fig.~\ref{fig:MW2Caroli} the illustrative diagrams for the process of derivation from Meir-Wingreen formula to Caroli formula.
It can be shown (see Appendix~\ref{app:compare2Mahan}) that the formula given by Eq.~(\ref{eq:Caroli}) and (\ref{eq:Trans}) is consistent with Mahan's and Yu's results\cite{Mahan2017, Yu2017}. The transmission function in Eq.~(\ref{eq:Trans}) can be written as
\begin{equation}
S = \frac{4 e^{-2 q d} \textrm{Im} (1 - 1/\varepsilon_{\bm{1}}^{\textrm{RPA}})  \textrm{Im}(1 -  1/\varepsilon_{\bm{2}}^{\textrm{RPA}}) }{|1 - e^{-2 q d} (1 - 1/\varepsilon_{\bm{1}}^{\textrm{RPA}})(1 -  1/\varepsilon_{\bm{2}}^{\textrm{RPA}}  )|^2}.
\end{equation}
This is also consistent with the evanescent part given by the theory of fluctuational electrodynamics~\cite{Volokitin_RMP2007, PhysRevB.85.155422}.

\section{\label{Calculation} Numerical results and discussion}
In the numerical calculation, we use one interacting layer for the central region on each side of the vacuum gap. We use 64 $\times$ 64 $\bm{k}$ points for the first Brillouin zone (FBZ). Fast fourier transformation (FFT) is used for calculating both the convolution in the energy space and the $\bm{k}$ space in the FBZ for the interacting self-energy to save computing time. Chemical potentials of the two leads are chosen to be $0 \; \textrm{eV}$. The lattice constant is chosen to be that of the gold~\cite{MILLER2006222} as $a=0.288 \; \textrm{nm}$. The hopping parameter in the cubic lattices is set to be $t=0.85 \; \textrm{eV}$. Energy range is chosen to cover the entire range of the  electron's spectrum. An energy cutoff of $2 \; \textrm{eV}$ is chosen for the scalar photon's GF.

\begin{figure}
\centering
\includegraphics[width=8.5 cm]{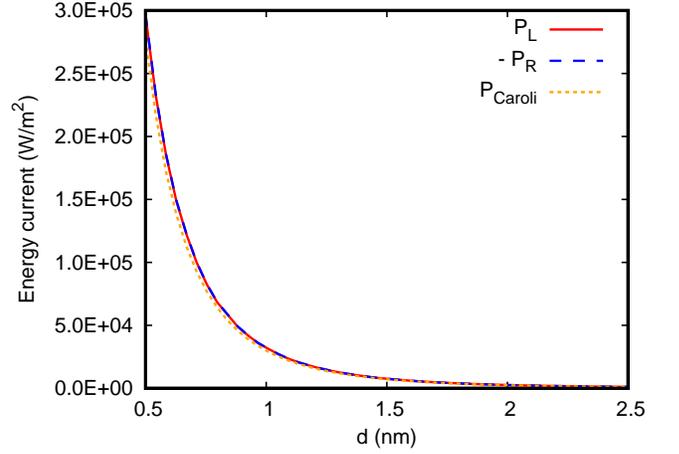}
\caption{Energy current by the Meir-Wingreen formula in the $G_{0}W_{0}$ approximation and by the Caroli formula. Electron tunneling is not included here, i.e., $t_{\textrm{gap}}=0$, $T_{L}=350 \; \textrm{K}$ and $T_{R}=300 \; \textrm{K}$. The corresponding value of black-body limit is $392 \; \textrm{W} / \textrm{m}^2$.}
\label{fig:PLPcaroli}
\end{figure}

Firstly, we discuss the energy transfer without the electron's tunneling, i.e. $t_{\textrm{gap}}=0$. We show in Fig.~\ref{fig:PLPcaroli} the energy current calculated by the Meir-Wingreen formula in the $G_{0}W_{0}$ approximation by Eq.~(\ref{eq:MWformula}) and the Caroli formula by Eq.~(\ref{eq:Caroli}) and (\ref{eq:Trans}). It is shown that the value given by the former is very close to that by the latter, indicating the local equilibrium approximation is a good approximation, which is used in the derivation from the former to the latter. The energy flow out of the left electrode is almost equal to that into the right electrode, i.e. $P_{L} \approx - P_{R}$. We note that $G_{0}W_{0}$ approximation doesn't obey the energy conservation, and the difference between $P_{L}$ and $P_{R}$ could be obvious in molecule systems~\cite{PhysRevB.77.115333}. The energy current is about two or three orders of magnitude larger than that given by the black-body limit in the extreme near distance. This is in agreement with the theory of fluctuational electrodynamics and the experimental results~\cite{Kim2015, Song2016}.

\begin{figure}
\centering
\includegraphics[width=8.5 cm]{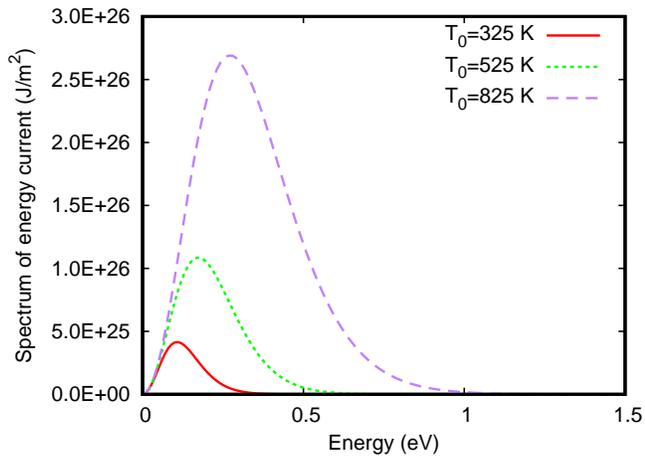}
\caption{Spectrum of the energy current by the Caroli formula. $T_{0}$ is the average temperature of the two electrodes, and $T_{L(R)}=T_{0}\pm \Delta T/2$, with $\Delta T=50 \; \textrm{K}$. The gap distance is $d = 0.6 \; \textrm{nm}$. $t_{\textrm{gap}}=0$.}
\label{fig:SpectrumEcurrent}
\end{figure}

We show in Fig.~\ref{fig:SpectrumEcurrent} the spectrum of the energy current, which is defined by $\frac{1}{A}\hbar \omega [N_{B}^{L}(\omega) - N_{B}^{R}(\omega)]   S(\omega)$ according to Eq.~(\ref{eq:Caroli}). Similar to the spectral radiance of the black-body radiation, with the increasing of the temperature, the peak of the spectrum of energy current moves to the higher energy part and the strength of the spectrum gets larger. The spectrum fades to 0 within an energy range of $1.5 \; \textrm{eV}$, which is much smaller than the typical energy scale of the plasmas in a good metal.

Secondly, we discuss the energy transfer including the electron's tunneling through the gap. We use the Simmons approximation~\cite{Simmons1963, PhysRevLett.27.922} to get an estimation for the distance dependence of the parameter $t_{\textrm{gap}}$. The Simmons approximation gives the relation between the electron's tunneling current and the gap distance in the low-voltage range as
$J = \frac{\sqrt{2 m_{e}}}{d} (\frac{e}{h})^2 U_{0} V e^{- C_{1} \sqrt{U_{0}}}$,
with $C_{1}=4\pi \beta d \sqrt{2 m_{e}}/h$. Here, $d$ is the gap distance, $m_{e}$ is the electron's mass, $h$ is the Planck constant, $V$ is the voltage, $U_{0}$ is the average potential between the metal surfaces, $\beta \approx 1$ is a constant. We use the approximation $J \propto t_{\textrm{gap}}^2$ to get the distance dependence of $t_{\textrm{gap}}$. Using the initial condition that $t_{\textrm{gap}}=t$  when the gap distance is equal to the lattice constant, we get the relation $t_{\textrm{gap}}=\sqrt{\frac{a}{d}} t e^{-C_{2} (d-a)/a}$, with $C_{2}=2 \pi \beta a \sqrt{2 m_{e}U_{0}}/h$.

\begin{figure}
\centering
\includegraphics[width=8.5 cm]{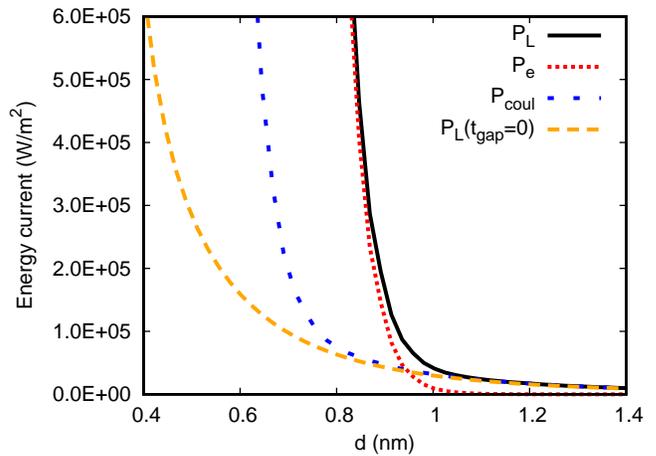}
\caption{Energy current including electron's tunneling. $T_{L}=350 \; \textrm{K}$, $T_{R}=300 \; \textrm{K}$. The average potential for the gap is $U_{0}=5.0 \; \textrm{eV}$.}
\label{fig:Ptgap}
\end{figure}

\begin{figure}
\centering
\includegraphics[width=8.5 cm]{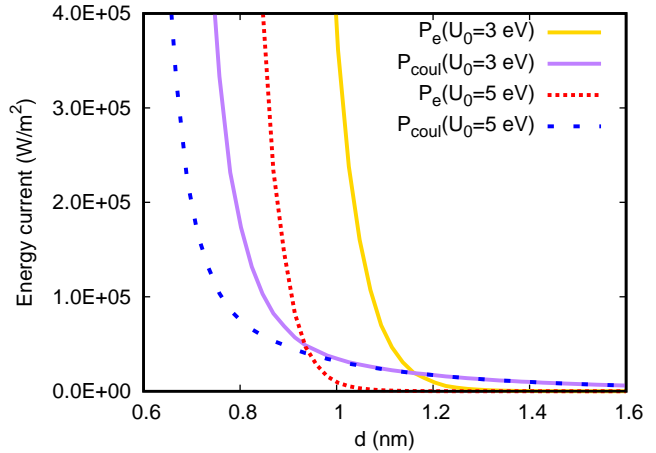}
\caption{Energy current including electron's tunneling with different average potentials for the gap. Other parameters are the same as those in Fig.~\ref{fig:Ptgap}.}
\label{fig:DiffU0}
\end{figure}

We show in Fig.~\ref{fig:Ptgap} the energy current including the electron's tunneling. The total energy current flowing out of the left electrode $P_{L}$ is given by Eq.~(\ref{eq:MWformula}). In order to get the separate contribution from electron's tunneling or Coulomb fluctuation by Eq.~(\ref{eq:MWformula}), denoted as $P_{\textrm{e}}$ or $P_{\textrm{coul}}$, we just need to replace $\bm{\Sigma}_{\textrm{tot}}^{<(>)}$ by $\bm{\Sigma}_{\textrm{leads}}^{<(>)}$ or  $\bm{\Sigma}^{F,<(>)}$ in the Keldysh equation for getting $\bm{G}^{<(>)}$.
With the gap distance varying from $0.4 \; \textrm{nm}$ to $1.4 \; \textrm{nm}$, the dominant contribution to energy current changes from the processes of electron's tunneling to the Coulomb fluctuation. The former changes much faster than the latter with the varying of the gap distance. The cross point is at a gap distance of about $0.92 \; \textrm{nm}$. When the gap distance is smaller than that of the cross point, the Coulomb contribution is much larger than that without electron's tunneling, implying that the electron's tunneling can enhance the Coulomb fluctuation in the strong tunneling region. Experiments show that when the vacuum gap is contaminated, the barrier potential could become smaller, and this can lead to larger energy current~\cite{Cui_NatCom2017}. In Fig.~\ref{fig:DiffU0}, we show the energy current by electron's tunneling and Coulomb fluctuation with different average potentials. When the average potential varies from $5 \; \textrm{eV}$ to $3 \; \textrm{eV}$, the cross point moves from about $0.92 \; \textrm{nm}$ to $1.15 \; \textrm{nm}$. This is consistent with the experimental case. However, the energy current decreases more quickly than the experimental case for a small barrier potential by the contaminants. A conducting model may be more appropriate, such as bridging the two metal plates by some quantum dots in the central region.

The barrier tunneling model we considered here has been widely used to consider electronic and energy transport through a single molecular junction.
Wherein, the height of the barrier is determined by the relative position of the molecular LUMO or HOMO orbital with respect to the electrode Fermi level.
Thus far, theoretical works studying energy transport through molecular junctions or atomic contacts use mainly noninteracting electron approximations,
for example, in the study of thermoelectric transport. Here, our results show that Coulomb fluctuation due to electron-electron interaction can modify the energy transport properties drastically. This has been largely overlooked before and needs to be  considered properly.

\section{\label{sec:Conclusion} Conclusion}
In summary, based on a tight-binding model, we study the energy transfer through the vacuum gap by the Coulomb fluctuation and electron tunneling between two cubic lattices using the NEGF method in the $G_{0}W_{0}$ approximation. Firstly, we get the Caroli formula from the Meir-Wingreen formula in the $G_{0}W_{0}$ approximation and local equilibrium approximation without including electron's tunneling. The Caroli formula is also consistent with the evanescent part given by the conventional theory of fluctuational electrodynamics, which plays an important role in the extreme near distance. Secondly, we go beyond the local equilibrium approximation to study the comparative energy transfer by Coulomb fluctuation and electron's tunneling. We focus on the crossover region for energy transfer from conducting limit to Coulomb fluctuation region using the Simmons approximation. We find that the Coulomb fluctuation is enhanced drastically in the strong tunneling region compared with the case of not including electron's tunneling. We compare our result with the experimental case with different barrier potentials, and we find that the value of the energy current by our tunneling model decreases much more quickly with the increasing of the gap distance compared with the experimental case for the small barrier potential due to contaminants, for which a conducting process or some other processes may be more appropriate.

\begin{acknowledgments}
J.T.L. is the support by the National Natural Science Foundation of China (Grant No. 61371015). J.S.W. is supported by FRC Grant No. R-144-000-343-112.
\end{acknowledgments}

\appendix

\section{Comparing Caroli formula with Mahan's result} \label{app:compare2Mahan}
We show that the Caroli formula is consistent with Mahan's result\cite{Mahan2017}. Here we focus on the distance dependence in the $z$ direction across the vacuum gap, neglecting this dependence inside the left or right interacting layers, so that we may make the approximation that the matrices of position indices in the $z$ direction such as $\bm{\Pi}_{\bm{1}\bm{1}}^{0,r}(\omega)$ and $\bm{D}_{\bm{1}\bm{2}}^{0,r}(\omega)$ are treated simply as numbers, rather than matrices. Notations are changed correspondingly, for example $\bm{\Pi}_{\bm{1}\bm{1}}^{0,r}(\omega) \rightarrow \Pi_{\bm{1}}^{0,r}(\omega)$. For the scalar photon's GF, we get from the Dyson equation in the RPA the relation $\bm{D}^{r} = [(\bm{D}^{0,r})^{-1} - \bm{\Pi}^{0,r}]^{-1}$, or in matrix form
\begin{equation}
\bm{D}^{r} = \left[ v_{q}^{-1} \left( \begin{array}{cc}
  1 &  e^{-q d}  \\
 e^{-q d} & 1
  \end{array}  \right)^{-1}
  -  \left( \begin{array}{cc}
   \Pi_{\bm{1}}^{0,r} & 0  \\
   0 & \Pi_{\bm{2}}^{0,r}
\end{array} \right)
\right]^{-1},
\end{equation}
with $v_{q} = \frac{1}{2 s_{0} \epsilon_{0} q}$. A direct calculation gives
\begin{equation}   \label{eq:D12r}
D_{\bm{1}\bm{2}}^{r} = \frac{v_{q} e^{-q d}}{1 - (v_{q} \Pi_{\bm{1}}^{0,r} + v_{q} \Pi_{\bm{2}}^{0,r}) + ( 1 - e^{-2 q d} ) v_{q} \Pi_{\bm{1}}^{0,r} v_{q} \Pi_{\bm{2}}^{0,r}}.
\end{equation}
Introducing the dielectric function $\varepsilon(\omega)$ and the density-density correlation function $\chi(\omega)$, which are related in the RPA by $\varepsilon^{\textrm{RPA}} = 1 - v_{q} \Pi^{0,r}$, $(\varepsilon^{\textrm{RPA}})^{-1} = 1 + v_{q} \chi^{\textrm{RPA}}$, $\chi^{\textrm{RPA}} = \Pi^{0,r}/(1 - v_{q} \Pi^{0,r})$, we can get the transmission function given in Eq.~(\ref{eq:Caroli}) as
\begin{equation}
S = \frac{4 e^{-2 q d} \textrm{Im} (1 - 1/\varepsilon_{\bm{1}}^{\textrm{RPA}})  \textrm{Im}(1 -  1/\varepsilon_{\bm{2}}^{\textrm{RPA}}) }{|1 - e^{-2 q d} (1 - 1/\varepsilon_{\bm{1}}^{\textrm{RPA}})(1 -  1/\varepsilon_{\bm{2}}^{\textrm{RPA}}  )|^2},
\end{equation}
where the relation $\textrm{Im}(\Pi^{0,r})/[\varepsilon^{\textrm{RPA}} (\varepsilon^{\textrm{RPA}})^\ast ]= \textrm{Im}(\chi^{\textrm{RPA}})$ is used. Using the notation $G_{L(R)}$ and $A_{L(R)}$, which are defined by $G_{L(R)} = 1 - 1/\varepsilon_{\bm{1}(\bm{2})}^{\textrm{RPA}}$ and $A_{L(R)} = -2 \textrm{Im}(1 - 1/\varepsilon_{\bm{1}(\bm{2})}^{\textrm{RPA}})$, and expanding $N_{B}^{L}(\omega, T_{0}+{\Delta T}/2)-N_{B}^{R}(\omega, T_{0}-{\Delta T}/2)$ with small quantity ${\Delta T}/T$, we can get from Eq.~(\ref{eq:Caroli}) the result
\begin{equation}
\begin{split}
P_{L} =& \frac{\Delta T}{ \hbar A k_{B} T_{0}^2} \int_{0}^{\infty} \frac{\textrm{d} \hbar \omega}{2 \pi} (\hbar \omega)^2 N_{B}(\omega, T_{0}) [1 + N_{B}(\omega, T_{0})] \\
& \times \sum_{\bm{q}} \frac{e^{-2qd} A_{L}(\bm{q}, \omega)A_{R}(\bm{q}, \omega)}{|1-e^{-2qd} G_{L}(\bm{q}, \omega) G_{R}(\bm{q}, \omega)|^2},
\end{split}
\end{equation}
which is consistent with the Eq. (31) by Mahan~\cite{Mahan2017}.

%

\end{document}